\documentclass[a4paper,12pt]{article}
\pdfoutput=1

\usepackage{jheppub}
\usepackage[utf8]{inputenc}
\usepackage[english]{babel}

\usepackage{amsfonts}
\usepackage{amsthm}
\usepackage{mathrsfs}
\usepackage[mathcal]{euscript}
\usepackage{float}
\usepackage[section]{placeins}

\newcommand{\be}{\begin{equation}}
\newcommand{\ee}{\end{equation}}

\newcommand{\skipline}{\vspace{\baselineskip}}

\DeclareMathOperator\arctanh{arctanh}
\newcommand{\nn}{\nonumber}
\newcommand{\eps}{\varepsilon}
\newcommand{\m}{_\text{m}}
\newcommand{\gen}{_\text{gen}}
\newcommand{\ext}{^\text{ext}}
\newcommand{\extgen}{\ext\gen}
\newcommand{\GN}{G_{\mathrm{N}}}
\newcommand{\LP}{\ell_\text{P}}
\newcommand{\tbreak}{t_\text{break}}
\newcommand{\LateTimes}{\Big\rvert_{\begin{subarray}{l}\text{late}\\\text{times}\end{subarray}}}
\newcommand{\EarlyTimes}{\Big\rvert_{\begin{subarray}{l}\text{early}\\\text{times}\end{subarray}}}
\newcommand{\ba}{\mathbf{a}}
\newcommand{\bb}{\mathbf{b}}
\newcommand{\bq}{\mathbf{q}}
\newcommand{\bp}{\mathbf{p}}
\newcommand{\bx}{\mathbf{x}}
\newcommand{\by}{\mathbf{y}}
\newcommand{\HH}{\mathcal{H}}
\newcommand{\const}{\text{const}}

\usepackage{comment}
\usepackage{soul}
\usepackage{xcolor}
\usepackage[normalem]{ulem}

\definecolor{darkblue}{rgb}{0,0,1}

\definecolor{dgreen}{rgb}{0,0.6,0}

\definecolor{darkraspberry}{rgb}{0.9,0.,0.3}

\definecolor{aquamarine}{rgb}{0.8,0.0,0.8}

\title{Entanglement entropy in de Sitter:\\ no pure states for conformal matter}

\author{D.S. Ageev,}
\author{I.Ya. Aref'eva,}
\author{A.I. Belokon,}
\author{V.V. Pushkarev}
\author{and T.A. Rusalev}

\affiliation{Steklov Mathematical Institute, Russian Academy of Sciences,\\ Gubkin str. 8, 119991 Moscow, Russia}

\emailAdd{ageev@mi-ras.ru}
\emailAdd{arefeva@mi-ras.ru}
\emailAdd{belokon@mi-ras.ru}
\emailAdd{pushkarev@mi-ras.ru}
\emailAdd{rusalev@mi-ras.ru}

\abstract{In this paper, we consider the entanglement entropy of conformal matter for finite and semi-infinite entangling regions, as well as the formation of entanglement islands in four-dimensional de Sitter spacetime partially reduced to two dimensions. We analyze complementarity and pure state condition of entanglement entropy of pure states as a consistency test of the CFT formulas in this geometrical setup, which has been previously used in the literature to study the information paradox in higher-dimensional de Sitter in the context of the island proposal. We consider two different types of Cauchy surfaces in the extended static patch and flat coordinates, correspondingly. For former, we found that entanglement entropy of a pure state is always bounded from below by a constant and never becomes zero, as required by quantum mechanics. In turn, the difference between the entropies for some region and its complement, which should be zero for a pure state, in direct calculations essentially depends on how the boundaries of these regions evolve with time. Regarding the flat coordinates, it is impossible to regularize spacelike infinity in a way that would be compatible with complementarity and pure state condition, as opposed, for instance, to two-sided Schwarzschild black hole. Finally, we discuss the information paradox in de Sitter and show that the island formula does not resolve it, at least in this setup. Namely, we give examples of a region with a time-limited growth of entanglement entropy, for which there is no island solution, and the region, for which entanglement entropy does not grow, but the island solution exists.}

\makeatletter
\gdef\@fpheader{\skipline}
\makeatother

\begin{document}
\maketitle
\flushbottom
\newpage


\section{Introduction}

De Sitter spacetime, which approximates our current Universe on large scales, as well as its inflationary stage in the first moments after the Big Bang, remains the most mysterious among all maximally symmetric spaces. A distinctive feature of de Sitter spacetime is the presence of causally disconnected regions. Each static observer can get access only to a part of the universe, contained within their cosmological horizon. This strongly distinguishes de Sitter spacetime from black holes, for which the horizon is not an observer-dependent notion. Due to this fact, the microscopic interpretation of thermodynamic entropy associated with the cosmological horizon seems to be different from that of black holes~\cite{Witten:2001kn, Spradlin:2001pw, Bousso:2002fq, Parikh:2002py, Goheer:2002vf, Parikh:2004ux, Parikh:2004wh, Anninos:2011af, Anninos:2012qw, Anninos:2017eib, Dong:2018cuv, Svesko:2022txo}. However, many properties of de Sitter and black holes are similar~\cite{PhysRevD.15.2738}. The cosmological horizon has the temperature proportional to its surface gravity and the coarse-grained entropy given by the Bekenstein-Hawking formula.

\skipline

It was previously believed that the entanglement entropy of an evaporating black hole exceeds the Bekenstein-Hawking entropy during time evolution. This contradicts the assumption that black hole evaporation is a unitary process, in which an initially pure state with vanishing entanglement entropy cannot evolve into a mixed state with non-zero entropy. However, under the assumption that black hole evaporation is unitary, Page showed~\cite{Page:1993wv, Page:2013dx} that despite the initial increase in the entanglement entropy of Hawking radiation, it eventually falls down to zero. This time evolution of the entanglement entropy of Hawking radiation is described by the so-called Page curve. The contradiction between direct calculations and the expected behaviour of the entanglement entropy, described by the Page curve, is one of the formulations of the information loss paradox.

Deriving the descending part of the Page curve is the main problem in the context of the information paradox, and recently the island proposal~\cite{Penington:2019npb, Almheiri:2019psf, Almheiri:2019hni} was introduced to do this, thus explaining how unitarity of black hole evaporation is to be restored. This conjecture suggests that entanglement entropy changes in the presence of dynamical gravity~\cite{Cotler:2017erl, Almheiri:2019psf}. Applying to the Hawking radiation of an evaporating black hole the so-called island formula, which was first derived in holography~\cite{Ryu:2006bv, Hubeny:2007xt, Barrella:2013wja, Faulkner:2013ana, Engelhardt:2014gca} and then using the replica trick~\cite{Almheiri:2019qdq, Penington:2019kki}, one finds that the fine-grained entropy follows the Page curve, initially increasing according to Hawking's calculations and then decreasing until it becomes zero~\cite{Penington:2019npb, Almheiri:2019psf, Almheiri:2019hni}. This approach has also been used to study the dynamics of entanglement entropy for eternal black holes~\cite{Almheiri:2019yqk, Almheiri:2019psy, Gautason:2020tmk, Anegawa:2020ezn, HIM, Hartman:2020swn, Dong:2020uxp, Kim:2021gzd}, for which the formulation of the information paradox is more obscure. The recent developments have been reviewed in~\cite{Almheiri:2020cfm, Marolf:2020rpm, Raju:2020smc}. Given the remarkable success of the island formula, one might wonder whether this progress can also shed light on the nature of the cosmological horizon.

\skipline

Even though the island formula has been derived only in two-dimensional JT gravity~\cite{Penington:2019kki, Almheiri:2019qdq, Almheiri:2019psf, Almheiri:2019hni}, where gravitational equations are solvable, it is believed that the main conclusions are the same in higher-dimensional gravity. Also, the analytical answer for entanglement entropy of matter is known only in some special models, including two-dimensional conformal field theory~\cite{Calabrese:2004eu, Casini:2005rm, Calabrese:2009qy}. More complicated setups, such as higher-dimensional spacetimes or spacetimes with no asymptotically flat subregions, are still subjects of research~\cite{HIM, Alishahiha:2020qza, Sybesma:2020fxg, Matsuo:2020ypv, Wang:2021woy, Kim:2021gzd, Aalsma:2021bit, Lu:2021gmv, Yu:2021cgi, Ahn:2021chg, Arefeva:2021kfx, Omidi:2021opl, Arefeva:2022cam, Gan:2022jay, Ageev:2022hqc, Djordjevic:2022qdk, Goswami:2022ylc, Yadav:2022jib, Yadav:2023qfg}. As for de Sitter, the literature uses a specific setup, in which the original manifold is either initially two-dimensional or is reduced to two dimensions, and the applicability of the island formula is postulated~\cite{Chen:2020tes, Hartman:2020khs, Balasubramanian:2020xqf, Sybesma:2020fxg, Aalsma:2021bit}.

\skipline

We study the properties of entanglement entropy of Hawking radiation in one of such setups, namely, four-dimensional de Sitter geometry, partially reduced to two dimensions. Similar geometry has been previously considered, for example, in~\cite{Sybesma:2020fxg,Svesko:2022txo, Aalsma:2021bit}. By considering Hawking quanta of free massless Dirac fermions collected in entangling regions of finite and semi-infinite extents, we reveal seemingly curious features of entanglement entropy in this geometry.

It is well known~\cite{Solodukhin:2011gn, Nishioka:2018khk} that a pure state has vanishing entanglement entropy
\be\nn
    S\,(\text{pure state}) = 0,
\ee
and, also, if this state is bipartite then the entropies of each partition are equal
\be\nn
    S\,(R) = S\left(\overline{R}\right).
\ee
In the following, we call these relations \emph{pure state condition} and \emph{complementarity}, respectively. The complementarity is commonly used in the calculations of entanglement entropy in a variety of setups, including semi-infinite regions in different black hole geometries~\cite{HIM, Alishahiha:2020qza, Matsuo:2020ypv, Wang:2021woy, Kim:2021gzd, Lu:2021gmv, Yu:2021cgi, Ahn:2021chg, Gan:2022jay, Goswami:2022ylc, Yadav:2022jib} and regions of finite extent in de Sitter spacetime~\cite{Sybesma:2020fxg}. However, we suggest calculating both sides of the last equality as well as checking the pure state condition to prove the applicability of the CFT techniques for calculating entanglement entropy on a given curved background.

Even though the geometries usually studied in the literature are conformally related to flat space (neglecting their angular parts), our suggestion is not as trivial as it might seem. As we have shown in the recent paper~\cite{Ageev:2022qxv}, pure state condition without a specific ``renormalization'' of the entropy is violated by a non-zero constant in two-sided Schwarzschild black hole. Also, one should use a special prescription for infrared regularization of infinite entangling regions in order for the complementarity property to hold.

\skipline

\begin{figure}[ht]\centering
    \includegraphics[width=0.4\textwidth]{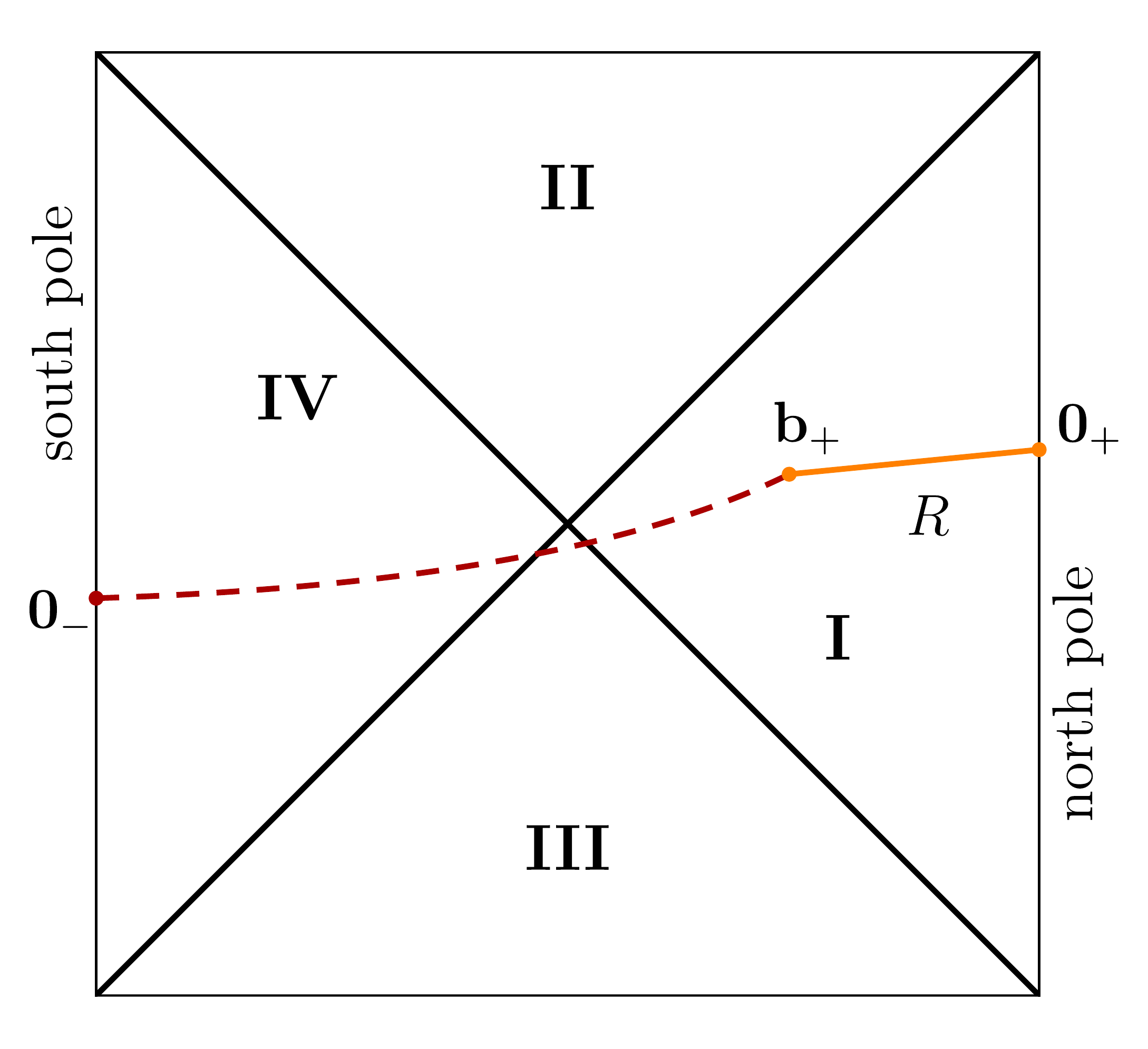} \hspace{0.05\textwidth}
    \includegraphics[width=0.4\textwidth]{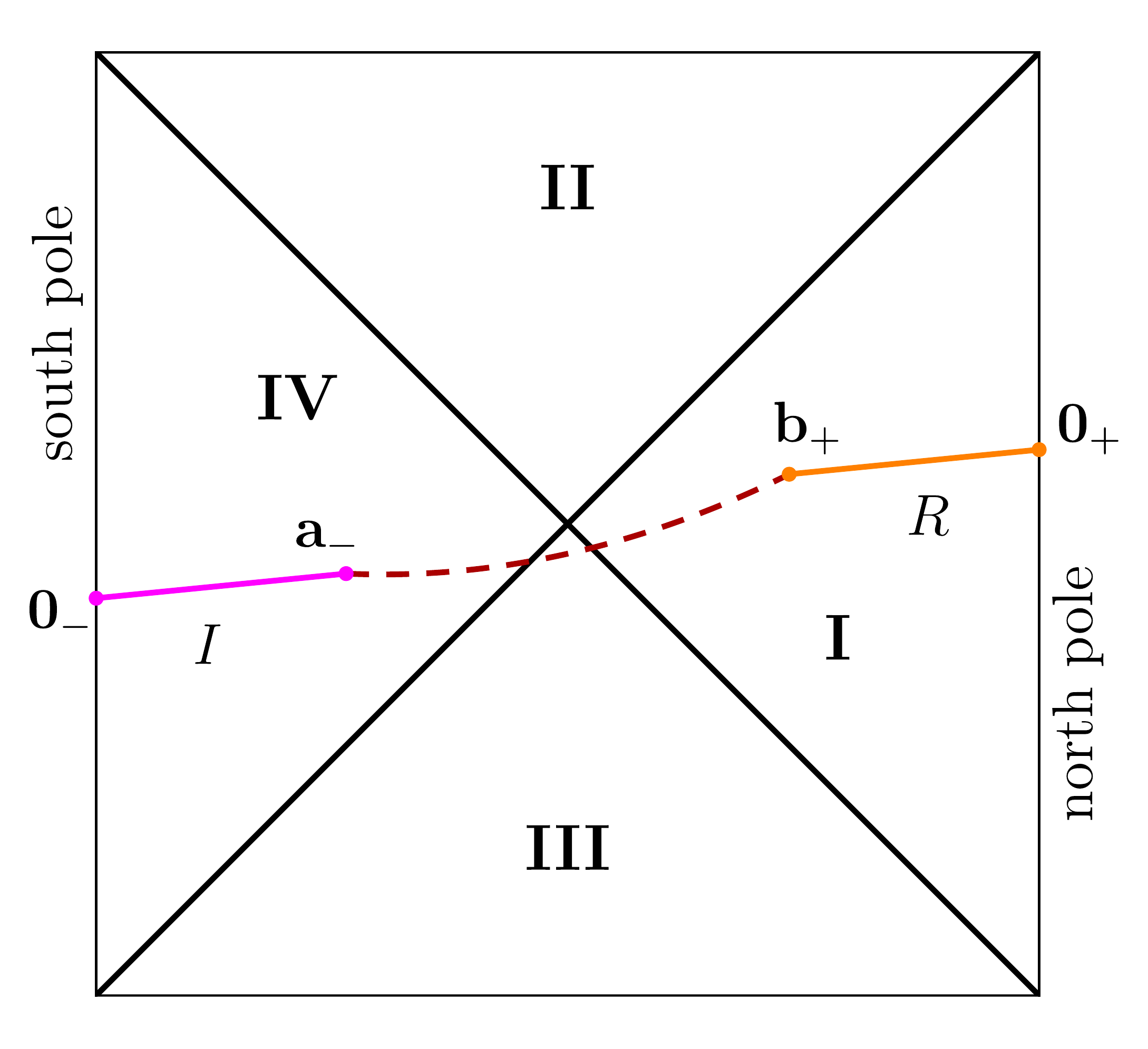}
	\caption{\emph{Left:} Penrose diagram for four-dimensional de Sitter spacetime with schematic plots of the entangling region $R = [\bb_+,\,\mathbf{0_+}]$ (orange) and its complement $\overline{R} = [\mathbf{0_-},\,\bb_+]$ (dashed dark red). \emph{Right:} the same but with the island $I = [\mathbf{0_-},\,\ba_-]$ (magenta) and the complement $\overline{R \cup I} = [\mathbf{0_-},\,\ba_-] \cup [\bb_+,\,\mathbf{0_+}]$ (dashed dark red).}
    \label{fig:LeftIslandScheme}
\end{figure}

An attempt to study a higher-dimensional de Sitter in the context of the island proposal was made in~\cite{Sybesma:2020fxg}, in which a finite entangling region (orange interval in figure~\ref{fig:LeftIslandScheme}) in the right wedge was considered. Using \emph{complementarity}, the entropy for this region was calculated as that for the complement one (dashed dark red interval in figure~\ref{fig:LeftIslandScheme}, left), which grows with time and exceeds the Gibbons-Hawking entropy of the cosmological horizon. To resolve the obtained information paradox, the island formula was applied and the entropy for the complement was used in the calculation (dashed dark red interval in figure~\ref{fig:LeftIslandScheme}, right). As a result, the author obtained the entanglement island moving back in time in the left wedge (magenta interval in~figure~\ref{fig:LeftIslandScheme},~right). This specific setup is interesting as an avatar of a higher-dimensional de Sitter, whose topology differs from that of the solution of the two-dimensional dilaton gravity with a positive cosmological constant, and also because there is a patch, in which the Cauchy surface is not infinitely large. The latter fact makes it possible to directly calculate both the total entanglement entropy of a pure state specified on such a Cauchy surface and the entropy of its individual subsystems without the need for infrared regularization of spacelike infinities. This allows us to check the basic properties of the entropy and draw a conclusion about the validity of this approach in resolving the information paradox in higher-dimensional de Sitter.

In this paper, we give an example of an entangling region, for which there is a time-limited monotonic growth but no island solution, so the possible information paradox cannot be resolved by the island formula. Also, we discuss a region, for which there is a non-trivial island configuration that never dominates the generalized entropy functional.

\skipline

The paper is organized as follows. In section~\ref{sec:setup}, we briefly review our setup. In section~\ref{sec:no-purity}, we explicitly check whether \emph{pure state condition} and \emph{complementarity} hold, and unveil that they are \emph{violated}, regardless of the type of Cauchy surfaces and their partitions into finite or semi-infinite regions. In section~\ref{sec:islands-and-inf-par}, we discuss whether there are the information paradox in de Sitter and possible applications of the island formula. We end with an outlook of our findings in section~\ref{sec:conclusions} and discuss them in the context of our previous results for the same problem in two-sided Schwarzschild black hole~\cite{Ageev:2022qxv}.


\section{Setup}
\label{sec:setup}
\subsection*{Geometry}

In this paper, two main coordinate charts of interest corresponding to the four-dimensional de Sitter spacetime are the static patch
\be
    ds^2 = - f(r) dt^{2} + \frac{dr^2}{f(r)} + r^2 d\Omega^2_{2}, \qquad f(r) = 1 - \frac{r^2}{\ell^2},
    \label{eq:StaticPatch}
\ee
and the flat patch
\be
    ds^2 = -d \tau^2 + e^{2\tau/ \ell} \left(d \rho^2 + \rho^2 d\Omega^2_{2}\right),
    \label{eq:FlatSlicing}
\ee
where $\ell$ is the de Sitter length, and $d\Omega^2_{2}$ is the angular part of the metric. The static patch~\eqref{eq:StaticPatch} covers only the right (or left) wedge of the Penrose diagram for the four-dimensional de Sitter spacetime, while the flat coordinates~\eqref{eq:FlatSlicing} cover its half.

The surface gravity $\kappa_c$ and the Hawking temperature $T$ associated with the cosmological horizon are given by
\be
    \kappa_c = \frac{|f^{'} (\ell)|}{2} = \frac{1}{\ell}, \qquad T = \frac{1}{2\pi\ell}. 
\ee

Kruskal coordinates covering the \emph{right wedge} of Penrose diagram can be expressed in terms of the static coordinates~\eqref{eq:StaticPatch} as follows
\be
    U = \frac{1}{\kappa_c}\,e^{\kappa_c(t - r_*(r))}, \quad V = -\frac{1}{\kappa_c}\,e^{-\kappa_c(t + r_*(r))},
    \label{eq:right_wedge_Krusk}
\ee
where the tortoise coordinate $r_{*}(r)$ is given by
\be
    r_{*}(r) = \frac{\ell}{2}\log\frac{\ell + r}{|\ell - r|} =
    \left\{\begin{array}{cc}
        \ell \arctanh{r/\ell}, & \quad r < \ell, \\
        \ell \arctanh{\ell/r}, & \quad r > \ell.
    \end{array}\right.
\ee
Having defined the static patch~\eqref{eq:StaticPatch}, we can obtain the Kruskal coordinates covering the \emph{left wedge} from those in the right~\eqref{eq:right_wedge_Krusk} with the following analytical continuation of the time coordinate $t$
\be
    t \to t + \frac{i\pi}{\kappa_c}.
    \label{eq:analyt-continue}
\ee
In the following, we will call the union of the left and right static patches related by this analytical continuation the \emph{extended static patch}.

In Kruskal coordinates~\eqref{eq:right_wedge_Krusk}, the metric takes the form
\be
    ds^2 = -e^{2\varrho(r)}dUdV + r^2d\Omega^2_{2},
    \label{eq:Kruslal-metric}
\ee
where the Weyl factor is given by
\be
    e^{2\varrho(r)} = f(r)e^{2\kappa_c r_{*}(r)} = \frac{4}{\left(1 - \kappa_c^2 U V\right)^2}.
    \label{eq:Weyl}
\ee

In what follows, we need a formula for the squared geodesic distance $d^2(\bx, \by)$ in the spherically symmetric two-dimensional part of the metric~\eqref{eq:Kruslal-metric}, which goes as
\be
    d^2(\bx, \by) = e^{\varrho(\bx)} e^{\varrho(\by)}\left[U(\bx) - U(\by)\right] \left[V(\by) - V(\bx)\right].
    \label{eq:d2}
\ee
Bold letters denote pairs of radial and time coordinates, e.g., $\bx = \left(x, t_x\right)$ or $\bx = \left(\rho_x, \tau_x\right)$.

In terms of the static coordinates~\eqref{eq:StaticPatch}, the distance takes the form
\be
    d^2(\bx, \by) = \frac{2 \sqrt{f(x) f(y)}}{\kappa^2_c} \big[\cosh \kappa_c (r_*(x) - r_*(y)) - \cosh\kappa_c (t_x - t_y)\big].
    \label{eq:DistStatic}
\ee
With~\eqref{eq:analyt-continue}, one can apply this formula to calculate the distance between the points in the extended static patch. We use the subscripts ``$+$'' and ``$-$'' to distinguish the points in the right and left static patches, respectively,
\be
    \bx_+ = \left(x_+,\,t_{x_+}\right), \quad \bx_- = \left(x_-,\,t_{x_-} + \frac{i \pi}{\kappa_c}\right).
\ee

The same distance rewritten in the flat coordinates~\eqref{eq:FlatSlicing} reads
\be
    d^2(\bx, \by) = \frac{1}{\kappa^2_c} \big[2 + \kappa^2_c (\rho_x - \rho_y)^2 e^{\kappa_c (\tau_x + \tau_y)} - 2\cosh \kappa_c (\tau_x - \tau_y)\big].
    \label{eq:DistPlanar}
\ee


\subsection*{Entanglement entropy and reduction to two dimensions}

We study entanglement entropy of $c$ copies of free massless Dirac fermions in the background of the four-dimensional de Sitter spacetime, using the extended static patch~\eqref{eq:Kruslal-metric} or flat coordinates~\eqref{eq:FlatSlicing}. The analytical calculation of the entanglement entropy in higher-dimensional setups ($d > 2$) is an extremely challenging problem. For this reason, various methods of reducing a higher-dimensional de Sitter to its two-dimensional counterpart are often used in the literature. It is worth emphasizing that the procedure of dimensional reduction of a pure (i.e. without matter) higher-dimensional de Sitter is in some sense ambiguous. In the spirit of~\cite{Aalsma:2021bit}, we distinguish between \emph{partial} and \emph{full} dimensional reductions of a higher-dimensional (four-dimensional in our case) de Sitter. When integrating out the angular part during the reduction to two dimensions, we obtain a dynamical field --- the dilaton, which is identified with the radial coordinate of the higher-dimensional metric~\cite{Achucarro:1993fd, Maxfield:2020ale, Maldacena:2019cbz, Cotler:2019nbi, Sybesma:2020fxg, Aalsma:2021bit}. The \emph{partial} reduction, in which we derive the avatar of the higher-dimensional de Sitter in two dimensions, imposes a restriction on the set of values of the dilaton, which must be positive, since it is essentially the absolute value of the radius-vector of the static observer. The \emph{full} reduction, which allows any dilaton values, gives a two-dimensional de Sitter, whose topology is different from the higher-dimensional~one. 

We are interested in the partially reduced gravitational action $\mathcal{S}_\text{\text{partially reduced dS$_2$}}$, to which the action of two-dimensional conformal matter represented by $c$ free massless Dirac fermions $\mathcal{S}_\text{CFT$_2$ fermions}$ is added, i.e.
\be
    \mathcal{S}_\text{pure dS$_4$} \to \mathcal{S}_\text{\text{partially reduced dS$_2$}} + \mathcal{S}_\text{CFT$_2$ fermions}.
\ee
This procedure leads to the following formula for the entropy of conformal matter for one interval~\cite{Sybesma:2020fxg}
\be
    S_\text{m} = \frac{c}{6}\log\frac{d^2(\bx, \by)}{\eps^2},
    \label{eq:S_1_ints}
\ee
where $d^2(\bx, \by)$ in our case is given by~\eqref{eq:DistStatic} or~\eqref{eq:DistPlanar}, and $\eps$ is a UV cutoff. For a system of $N$ intervals~\cite{Casini:2005rm}, it is given by
\be
    S\m = \frac{c}{6} \sum_{i,\,j} \log\frac{d^2(\bx_{i}, \by_{j})}{\eps^2} - \frac{c}{6} \sum_{i\,<\,j} \log\frac{d^2(\bx_{i}, \bx_{j})}{\eps^2} - \frac{c}{6} \sum_{i\,<\,j} \log\frac{d^2(\by_{i}, \by_{j})}{\eps^2},
    \label{eq:S_N_ints}
\ee
where $\bx_i$ and $\by_i$ denote the left and right endpoints of the corresponding intervals. Some of our results, which are based only on~\eqref{eq:S_1_ints}, are valid for any CFT$_2$, not necessarily for fermions.

\skipline

\begin{figure}[t!]\centering
    \includegraphics[width=0.65\textwidth]{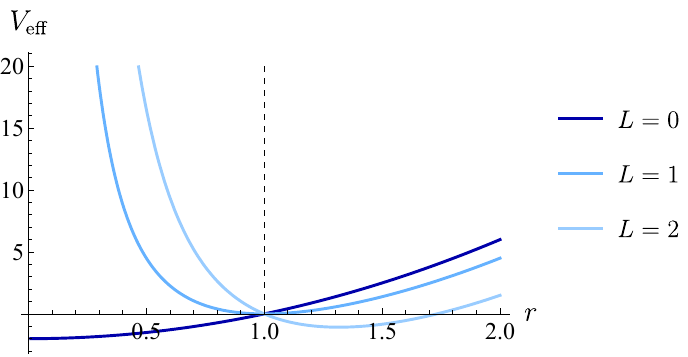}
    \caption{Effective potential~$V_{\text{eff}}$~\eqref{eq:Veff} of the wave equation in the static patch~\eqref{eq:StaticPatch} of de Sitter as a function of the distance from the static observer for several lowest multipoles, $L = 0$, $L = 1$ and $L = 2$. The cosmological horizon is at $\ell = 1$ and is depicted in dashed black. The zeroth multipole (s-wave) is strongly distinguished: $V_{\text{eff}}$ under the horizon for $L = 0$ has the form of a potential well, while for higher multipoles $L > 0$ in the bulk $r < \ell$ it acts as a potential barrier that scatters the radiation back to the horizon. The higher the multipole, the stronger the backscattering.}
    \label{fig:app:VEff}
\end{figure}

Application of these formulas can also be motivated by the s-wave approximation~\cite{Penington:2019npb, HIM}, whereby we neglect the spherical part of the metric~\eqref{eq:Kruslal-metric} and effectively reduce the initial setup to a two-dimensional CFT problem, in which the entanglement entropy can be calculated analytically. The s-wave approximation for Schwarzschild spacetime in the context of calculating the entanglement entropy was first applied in~\cite{HIM}. The effective reduction to a two-dimensional problem is justified by the analysis of backscattering of higher multipoles by the black hole effective potential. Since the potential barrier is the smallest for the lowest multipole (s-wave), the spherically symmetric part of the radiation dominates at spacelike infinity, where the static observer collects Hawking quanta. The same analysis can be done for the wave equation in the static patch~\eqref{eq:StaticPatch} of de Sitter, whose effective potential is given by (see figure~\ref{fig:app:VEff})
\be
    V_{\text{eff}}(r) = f(r)\left(\frac{L(L + 1)}{r^2} -\frac{2}{\ell^2}\right),
    \label{eq:Veff}
\ee
where $L$ is the number of a multipole. The applicability of the s-wave approximation is now determined by the behaviour of this potential at the origin $r = 0$, where the static observer collects Hawking radiation coming from the cosmological horizon. 

In this paper, we will not elaborate on a specific procedure for reducing to two dimensions. Instead, we will take this partially reduced geometry as the starting point. To study the properties of entanglement entropy of Hawking radiation in de Sitter spacetime, we will use the formulas~\eqref{eq:S_1_ints} and~\eqref{eq:S_N_ints}. Notice that, as discussed in \cite{Svesko:2022txo, Aalsma:2021bit}, in the partially reduced geometry of three-dimensional de Sitter, the constant-time hypersurface is a semicircle with endpoints at the north and south poles. In \cite{Svesko:2022txo} it is suggested that, probably, it would be more natural to consider quantum field theory in the presence of boundaries. In this paper, we consider conformal field theory without boundaries in the manner of~\cite{Svesko:2022txo, Aalsma:2021bit}. As we have pointed out previously, one can consider our calculations as a consistency test of such a setup.


\subsection*{Generalized entropy functional}

The island proposal is currently seen as a prime candidate for resolving the information loss paradox~\cite{Cotler:2017erl, Almheiri:2019hni, Penington:2019npb, Almheiri:2019qdq}. When computing the fine-grained entropy in gravitational theories using the replica trick, the contribution of some region, called the entanglement island, appears as a new saddle to the Euclidean semi-classical gravitational path integral, thus changing the dynamics of the entropy. When the entanglement between the gravitating region and the region where there is no gravity (this might be either an external bath or an asymptotically flat part of the spacetime itself) becomes strong, this saddle dominates, making the entropy evolution unitary.

For four- and higher-dimensional Schwarzchild black hole, it was shown in \cite{HIM} that, in some approximation, such a prescription recovers the expected behaviour of the Page curve. Considering the Hartle-Hawking vacuum~\cite{Hartle:1983ai}, the reduced density matrix of Hawking radiation, collected in an entangling region~$R$, is defined by tracing out the states in the complement region~$\overline{R}$, which includes the black hole interior. The island proposal prescribes that the states in the entanglement island~${I \subset \overline{R}}$ are to be excluded from tracing out. 

Despite being successful in resolving the information paradox for black holes, there are issues with the island proposal in de Sitter spacetime. Indeed, there is no asymptotically flat subregion, where an observer can collect Hawking radiation coming from the cosmological horizon, to which the island formula can be applied. However, we postulate the applicability of the island proposal in de Sitter spacetime, as it is often done in the literature~\cite{Hartman:2020khs, Sybesma:2020fxg, Aalsma:2021bit, Chen:2020tes, Balasubramanian:2020coy, Balasubramanian:2020xqf, Geng:2021wcq}.

The main conjecture (in higher-dimensional case) of the island proposal is that the formula for entanglement entropy in the presence of dynamical gravity gets modified. The island contribution is taken into account via the extremization of the generalized entropy functional~\cite{Penington:2019kki, Almheiri:2019qdq}, which is defined~as
\be
    S\gen[I, R] = \frac{\operatorname{Area}(\partial I)}{4\GN} + S\m(R \cup I).
    \label{eq:gen_functional}
\ee
Here $\partial I$ denotes the boundary of the entanglement island, $\GN$ is Newton's constant, and $S\m$ is the entanglement entropy of matter (in our case defined by~\eqref{eq:S_1_ints} and~\eqref{eq:S_N_ints}). One should extremize this functional over all possible island configurations
\be
    S\extgen[I, R] = \underset{\partial I}{\operatorname{ext}}\,\Big\{S\gen[I, R]\Big\},
\ee
and then choose the minimal one
\be
    S(R) = \underset{\partial I}{\text{min}}\,\Big\{S\extgen[I, R]\Big\}.
    \label{eq:Sgen}
\ee


\section{No pure state condition and complementarity in de Sitter}
\label{sec:no-purity}

Let us consider a pure bipartite quantum state, whose Hilbert space is factorized into two parts according to the partition of the Cauchy surface, ${\Sigma = R \cup \overline{R}}$ and ${\HH_\Sigma = \HH_R \otimes \HH_{\overline{R}}}$. For this state, the entanglement entropy of the entire Cauchy surface is zero, while the entanglement entropies of each partition are equal to each other~\cite{Solodukhin:2011gn, Nishioka:2018khk}
\be
    S(\Sigma) = 0, \quad S(R) = S\left(\overline{R}\right).
    \label{eq:ComplProp}
\ee
We call these relations \emph{pure state condition} and \emph{complementarity}, respectively.

Since we are dealing with a theory \emph{in the presence of gravity}, it can be argued that in order to infer whether complementarity holds, we should compare the entanglement entropies \emph{given by the island formula}~\eqref{eq:Sgen}. However, if one of the following statements is true for both the region and its complement, namely:

a) the islands are the same for both the entangling region and its complement;

b) there are no islands in the setup;

c) there are islands, but they never dominate the generalized entropy functional;\\
then it is sufficient to compare the entanglement entropies of matter only. In this paper, we assume that one of these conditions is always met and provide a few supporting arguments in section~\ref{sec:islands-and-inf-par}.

In this section, we show via direct calculations that pure state condition is violated in the extended static patch of de Sitter, as well as that the entropies in the definition of the complementarity property~\eqref{eq:ComplProp} are essentially different, so there is no way to regularize this difference and obtain the same expressions for $S\m(R)$ and $S\m\left(\overline{R}\right)$ neither in the extended static patch nor in the flat coordinates. Note that the results of this section are valid for any conformal field theory in two dimensions and any simply connected entangling region.


\subsection{Extended static patch}

\subsubsection*{Entanglement entropy for Cauchy surface}

\begin{figure}[ht]\centering
    \includegraphics[width=0.7\textwidth]{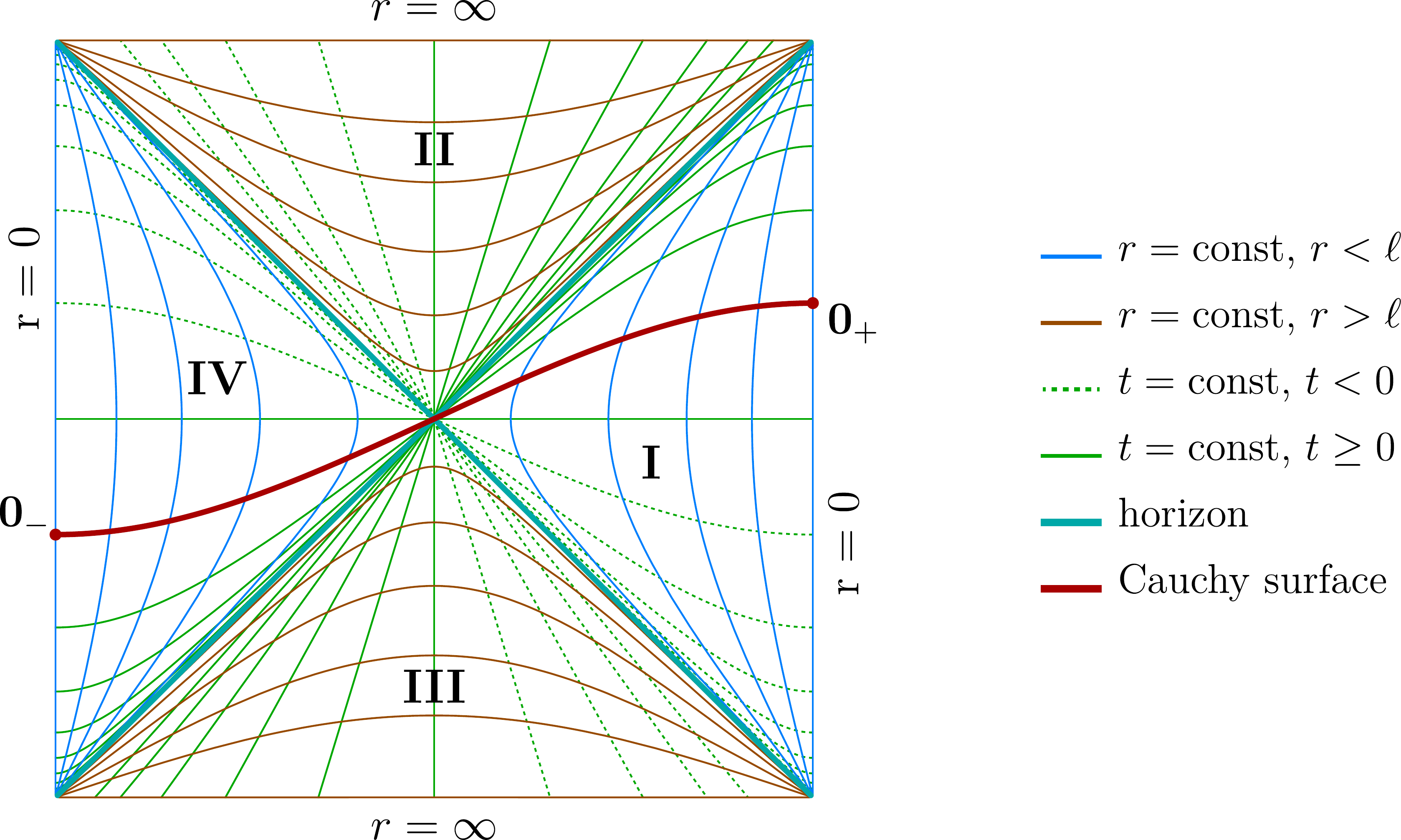}
	\caption{Penrose diagram for the extended static patch of the four-dimensional de Sitter spacetime with a Cauchy surface (dark red) stretched between the timelike hypersurfaces $r = 0$ in the left and right wedges: $\Sigma = [\mathbf{0_-},\,\mathbf{0_+}]$.}
    \label{fig:CauchySurfStatic}
\end{figure}

Cauchy surfaces of the extended static patch of de Sitter are stretched between the timelike hypersurfaces $r = 0$ in the right and left wedges: $\Sigma \equiv [\mathbf{0_-}, \mathbf{0_+}]$, see figure~\ref{fig:CauchySurfStatic}, with the points $\mathbf{0_\pm}$ defined as
\be
    \mathbf{0_+} = \left(0,\,t_{0_+}\right), \quad \mathbf{0_-} = \left(0,\,t_{0_-} + \frac{i \pi}{\kappa_c}\right).
\ee

Using the expression for the entanglement entropy~\eqref{eq:S_N_ints} along with~\eqref{eq:DistStatic} we obtain the entanglement entropy associated with the Cauchy surface $\Sigma$
\be
    S\m(\Sigma) = \frac{c}{6}\log\frac{d^2(\mathbf{0_+}, \mathbf{0_-})}{\eps^2} = \frac{c}{6}\log\left(\frac{2}{\kappa^2_c \eps^2} \left[1 + \cosh \kappa_c (t_{0_+} - t_{0_-}) \right]\right).
    \label{eq:EntrCauchySurf}
\ee
We conclude that there is the lower bound on the entropy for Cauchy surfaces
\be
    S\m(\Sigma) \geq \frac{c}{3}\log\frac{2}{\kappa_c \eps}.
    \label{eq:EntrCauchySurf_lowerbound}
\ee
This lower bound is achieved when $t_{0_+} = t_{0_-}$ and in particular, this also holds for the Cauchy surfaces, which are hypersurfaces of constant time in the extended static patch.

This result is in contradiction with pure state condition, which says that $S_\text{m}(\Sigma) = 0$. Formally, we could make the entropy for a Cauchy surface equal to zero by imposing the following relation
\be
    t_{0_-} = t_{0_+} + \frac{i \pi}{\kappa_c}.
    \label{eq:ZeroesRelation}
\ee
This implies the topology of a one-dimensional sphere $S^1$ for Cauchy surfaces in the extended static patch. Since the entropy in CFT$_2$ depends only on the endpoints of an interval, see~\eqref{eq:S_1_ints}, complementarity on $S^1$ holds automatically, because the endpoints of an entangling region and its complement coincide (see also section III in~\cite{Ageev:2022qxv}). However, such identification of the world lines of the origins $r = 0$ in the left and the right wedges contradicts the geometry of higher-dimensional de Sitter spacetime, in which they are opposite poles of the Penrose diagram, and hence, take different values of the polar angle.


\subsubsection*{Entanglement entropy for finite intervals}

\begin{figure}[ht]\centering
    \includegraphics[width=0.7\textwidth]{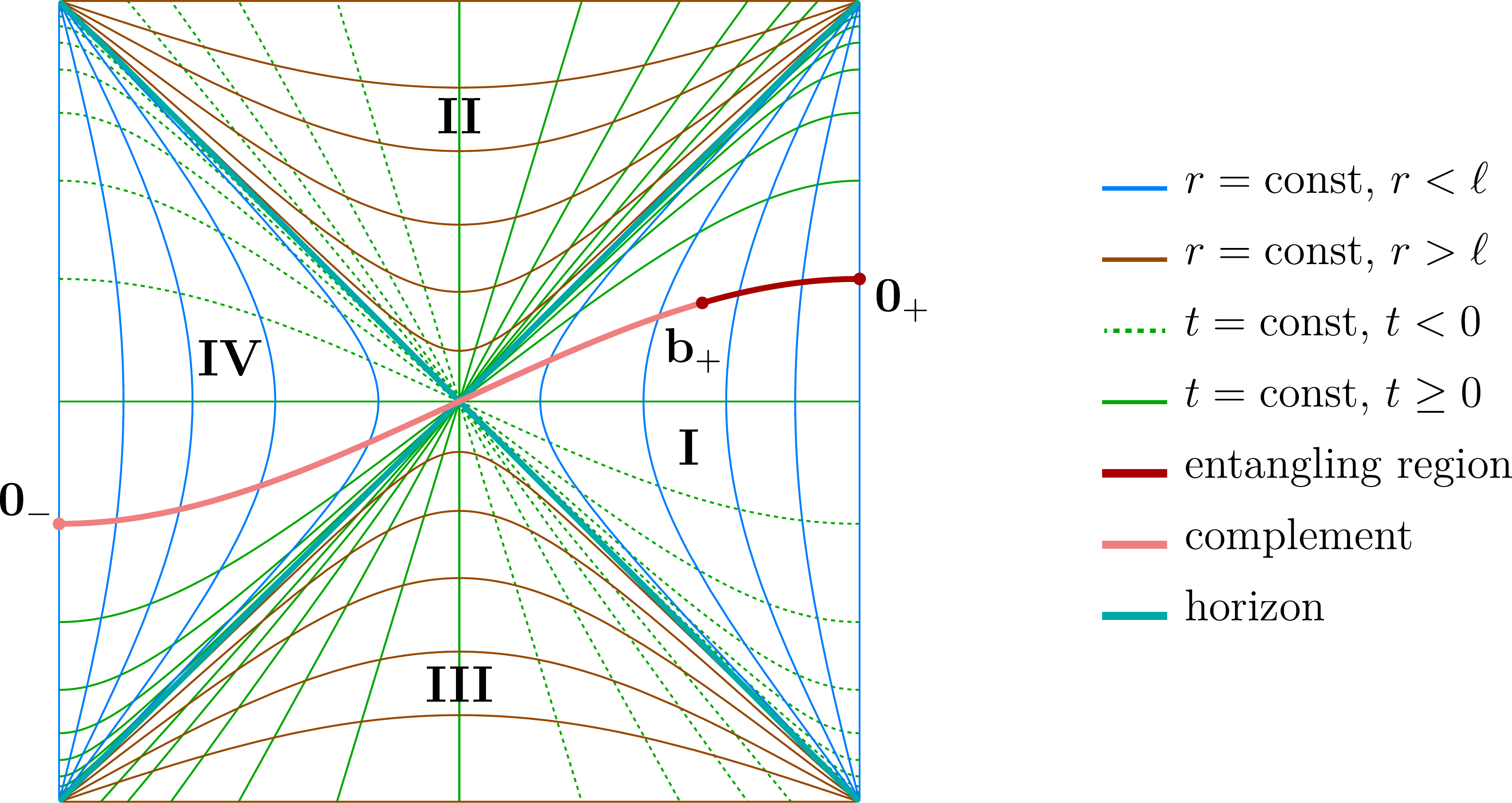}
	\caption{Penrose diagram for the extended static patch of the four-dimensional de Sitter spacetime with an entangling region $[\bb_+,\,\mathbf{0_+}]$ (dark red) and its complement $[\mathbf{0_-},\,\bb_+]$ (light red).}
    \label{fig:ComplPropStatic}
\end{figure}

Let us consider the entanglement entropy for the simplest entangling region
\be
    R = [\bb_+,\,\mathbf{0_+}], \quad \bb_+ = \left(b,\,t_b\right), \quad \mathbf{0_+} = \left(0,\,t_{0_+}\right),
\ee 
located in the right wedge (see figure~\ref{fig:ComplPropStatic}). It is given by
\be
    S\m(R) = \frac{c}{6}\log\left(\frac{2\sqrt{f(b)}}{\kappa^2_c \eps^2} \left[\cosh \kappa_c r_{*} (b) - \cosh \kappa_c (t_b - t_{0_+})\right]\right).
    \label{eq:R-fin}
\ee 
The complement of this region, $\overline{R} = [\mathbf{0_-},\,\bb_+]$, spans between two wedges. Its entanglement entropy reads
\be
    S\m\left(\overline{R}\right) = \frac{c}{6}\log\left(\frac{2\sqrt{f(b)}}{\kappa^2_c \eps^2} \left[\cosh \kappa_c r_{*} (b) + \cosh \kappa_c (t_b - t_{0_-})\right]\right).
    \label{eq:R-fin-compl}
\ee
These expressions have different relative signs in the brackets, therefore the complementarity property for these regions is explicitly violated. This would hold only if we impose the identification of the origins~\eqref{eq:ZeroesRelation}, which, as we argued above, is geometrically prohibited.

\begin{figure}[ht]\centering
    \includegraphics[width=0.8\textwidth]{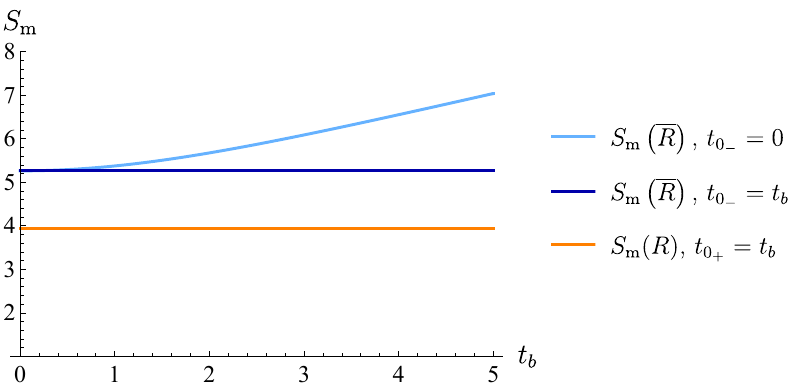}
    \caption{Time evolution of entanglement entropies $S\m(R)$ and $S\m\left(\overline{R}\right)$ for the region $R = [\bb_+,\,\mathbf{0_+}]$ and its complement $\overline{R} = [\mathbf{0_-},\,\bb_+]$ for different choices of the time coordinates of the endpoints. Other parameters are fixed as $b = 0.5$, $\ell = 1$, $c = 3$, $\eps = 0.01$. $S\m(R)$ and $S\m\left(\overline{R}\right)$ demonstrate essentially different dynamics depending on the choice of the time coordinates of the origins $r = 0$, thus violating the complementarity property.}
    \label{fig:FinRegDyn}
\end{figure}

The specific details of how the entropies differ (and hence, how the complementarity property is violated) are determined by the choice of the time coordinates $t_b$, $t_{0_+}$ and $t_{0_-}$, refer to figure~\ref{fig:FinRegDyn}. For instance, choosing all endpoints so that they lie on the same time slice, i.e. $t_b = t_{0_+} = t_{0_-}$, one can get that the entropies~\eqref{eq:R-fin} and~\eqref{eq:R-fin-compl} do not show any dynamics and hence, they differ only by a constant (see the blue and orange curves in figure~\ref{fig:FinRegDyn}). On the other hand, if one chooses the endpoint in the left wedge at some fixed time, say $t_{0_-} = 0$, then the behaviour changes drastically (see the light blue curve in figure~\ref{fig:FinRegDyn}). Indeed, the entropy of the region in the right wedge is still constant, while the entropy of its complement becomes time-dependent and grows linearly in $t_b$ at late times $\kappa_c t_b \gg 1$.

This particular situation is encountered in the setup discussed in~\cite{Sybesma:2020fxg}, which we have briefly described in the introduction. Assuming the complementarity property holds (which, as we have shown above, does not) and making the special choice $t_{ 0_-} = 0$, we obtain a monotonic growth of the entanglement entropy, which leads to the information loss paradox. If such a special choice is not made, then the entropy is constant, and no paradox arises.


\subsection{Flat coordinates}

Unlike the static patch, in the flat coordinates~\eqref{eq:FlatSlicing}, which describe an inflationary universe, the spacelike infinity $i^0$ is not closed from the observer by the cosmological horizon. If Cauchy surfaces contain $i^0$, as they do in the flat coordinates~\eqref{eq:FlatSlicing}, then it is necessary to regularize spacelike infinity~$i^0$ in order to calculate entanglement entropy given by~\eqref{eq:S_1_ints} and~\eqref{eq:S_N_ints}.

\subsubsection*{Entanglement entropy for Cauchy surface}

\begin{figure}[ht]\centering
    \includegraphics[width=0.85\textwidth]{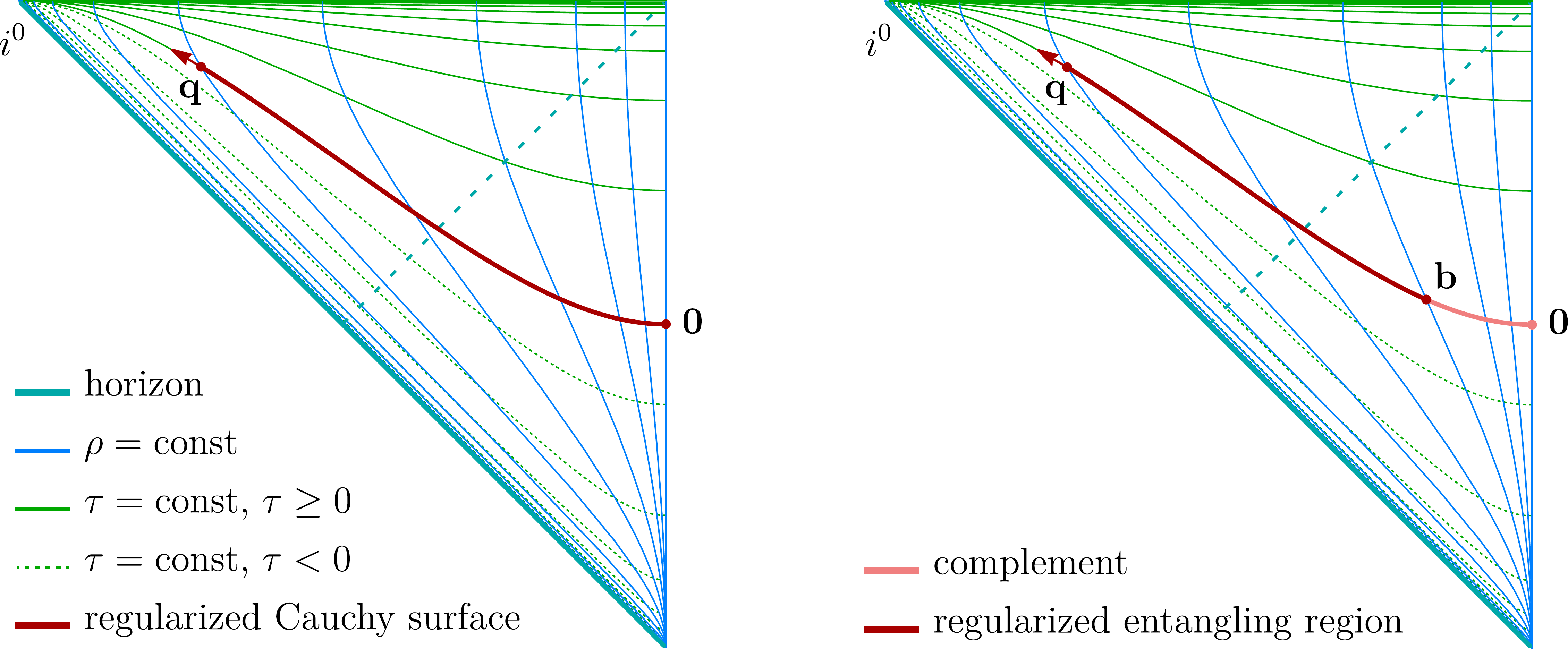}
    \caption{\emph{Left:} Penrose diagram for the flat coordinate patch of the four-dimensional de Sitter spacetime with a regularized Cauchy surface~$\Sigma_\text{reg} = [\bq,\,\mathbf{0}]$. \emph{Right:} the same with a regularized semi-infinite entangling region $[\bq,\,\bb]$ (dark red) and its complement $[\bb,\,\mathbf{0}]$ (light red). The arrow highlights that the regulator tends to spacelike infinity, $\bq \to i^0$.}
    \label{fig:CauchySurfPlanar}
\end{figure}

Cauchy surfaces $\Sigma$ in the flat patch~\eqref{eq:FlatSlicing} are extended from $\rho = 0$ to $\rho \to \infty$. To consider the entanglement entropy for (semi-)infinite regions we introduce a regulator~$\bq = \left(\rho_q,\,\tau_q\right)$ for spacelike infinity $i^0$ (see figure~\ref{fig:CauchySurfPlanar}, left).

The entanglement entropy for the regularized Cauchy surface $\Sigma_\text{reg} = [\mathbf{0}, \, \bq]$ is given by
\be
    S\m(\Sigma_\text{reg}) = \frac{c}{6}\log\frac{d^2(\bq,\,\mathbf{0})}{\eps^2} = \frac{c}{6}\log\left(\frac{\rho^2_q}{\eps^2}\,e^{\kappa_c(\tau_q + \tau_0)} +\frac{2}{\kappa^2_c\eps^2}\Big[1 - \cosh \kappa_c(\tau_q - \tau_0)\Big]\right).
    \label{eq:EntrCauchySurfPlanar}
\ee
Since $\Sigma_\text{reg} \to \Sigma$ as $\bq \to i^0$, we expect that $S(\Sigma_\text{reg}) \to S(\Sigma)$, which has to be~zero according to pure state condition. However, for any fixed values of the time coordinates~$\tau_q$ and $\tau_0$, the entropy becomes infinite in the limit $\rho_q \to\infty$
\be
    \lim_{\rho_q\,\to\,\infty} S\m(\Sigma_\text{reg}) \to \infty, \quad \tau_{q,\,0} = \const.
\ee

Let us discuss the possibility of introducing a specific regularization that would make it possible to satisfy pure state condition. Let $\tau_0 = 0$ and $\tau_q \neq \tau_0$. The leading-order expansion of regularized pure state condition $S\m(\Sigma_\text{reg}) = 0$ gives the relation
\be
    \tau_q(\rho_q) \approx -\frac{1}{\kappa_c}\log\kappa_c \rho_q + O(1/\rho_q).
\ee
Hence, to satisfy pure state condition, the regulator $\rho_q$ formally should be chosen as
\be
    \rho_q(\tau_q) = \frac{1}{\kappa_c}\,e^{-\kappa_c \tau_q}.
    \label{eq:reg_prescr}
\ee

Several comments are in order. To obtain the whole Cauchy surface, we should send the regulator $\bq$ to spacelike infinity $\rho_q \to \infty$ in the upper left edge of the Penrose diagram. According to~\eqref{eq:reg_prescr}, this implies that $\tau_q \to -\infty$. However, the metric \eqref{eq:FlatSlicing} has a coordinate singularity at $\tau \to -\infty$, since the scale factor $e^{\tau/\ell}$ becomes zero. The latter means that there is no spatial section with points at spacelike infinity. Therefore, the simultaneous limit $\rho_q \to \infty, \, \tau_q \to -\infty$, under which the entropy for the Cauchy surface~\eqref{eq:EntrCauchySurfPlanar} is finite, is not well-defined. For this reason, we write the relation~\eqref{eq:reg_prescr} only formally.


\subsubsection*{Entanglement entropy for semi-infinite interval}

To check whether we can move further in this direction, let us consider the partition of a Cauchy surface into a semi-infinite entangling region $[\bq,\,\bb]$ and its finite complement~$[\bb,\,\mathbf{0}]$ (see figure~\ref{fig:CauchySurfPlanar}, right). The entropy for the regularized region $[\bq,\,\bb]$~is given by
\be
    \begin{aligned}
        S\m\left([\bq,\,\bb]\right) & = \frac{c}{6} \log\frac{d^2(\bq,\,\bb)}{\eps^2} = \\
        & = \frac{c}{6} \log\left(\frac{(\rho_q - \rho_b)^2}{\eps^2} e^{\kappa_c (\tau_q + \tau_b)} +\frac{2}{\kappa^2_c \eps^2} \left[1 - \cosh \kappa_c (\tau_q - \tau_b)\right] \right).
    \end{aligned}
\ee
If $\tau_q = \const$, then this expression diverges at $\rho_q \to \infty$. For the special choice of the regulator~\eqref{eq:reg_prescr}, which would formally preserve pure state condition for a Cauchy surface, we obtain a finite limit
\be
    \lim_{\rho_q\,\to\,\infty} S\m\left([\bq,\,\bb]\right) = \frac{c}{6}\log\left(\frac{2 - 2\kappa_c \rho_b e^{\kappa_c \tau_b}}{\kappa_c^2\eps^2}\right).
\ee
However, it does not coincide with the entropy for the complement
\be
    S\m\left([\bb,\,\mathbf{0}]\right) = \frac{c}{6} \log\frac{d^2(\bb,\,\mathbf{0})}{\eps^2} = \frac{c}{6} \log\left(\frac{\rho^2_b}{\eps^2} e^{\kappa_c (\tau_b + \tau_0)} +\frac{2}{\kappa^2_c \eps^2} \left[1 - \cosh \kappa_c (\tau_b - \tau_0)\right]\right).
\ee
Hence, even with a special ``regularization''~\eqref{eq:reg_prescr}, the complementarity does not~hold.

\skipline

Since the flat patch~\eqref{eq:FlatSlicing} contains spacelike infinity $i^0$, in contrast to the extended static patch~\eqref{eq:StaticPatch}, let us compare the obtained results for entanglement entropy in de Sitter with those in two-sided Schwarzschild black hole, in which it is also necessary to introduce IR regulators for $i^0$, see~\cite{Ageev:2022qxv} for details.

Entanglement entropy diverges in the limit $\rho \to \infty$ due to the fact that, unlike two-sided Schwarzschild black hole, Cauchy surfaces in the flat patch contain only one spacelike infinity $i^0$, and hence, they are regularized by only one regulator $\bq$, see~\eqref{eq:EntrCauchySurfPlanar}. If we had two spacelike infinities $i^0$, then some kind of a symmetric regularization would be possible, in which two regulators $\bq_1$ and $\bq_2$ cancel each other at $\bq_{1,2} \to i^0$, giving a finite answer for the entropy. However, even the presence of two regulators is not sufficient. As it follows from the definition of the squared distance~\eqref{eq:d2}, another condition is the convergent behaviour of the Weyl factor at infinity. The Weyl factor in the flat coordinates~\eqref{eq:Weyl} is given by
\be
    e^{\varrho} = 1 + \kappa_c \rho e^{\kappa_c \tau},
\ee 
and leads to the following behaviour of the squared distance~\eqref{eq:DistPlanar} at $\bq \to i^0$
\be
    d^2(\bq,\,\textbf{0}) \approx \rho_q^2 e^{\kappa_c\tau_q} + O(1),
\ee
i.e. the squared distance diverges quadratically. This is opposed to two-sided Schwarzschild, in which there are two regulators $\bq_\pm$ in both left and right static patches, and the Weyl factor behaves as $e^{-2\kappa_h r}/r$, suppressing the divergence of the squared distance.


\section{Entanglement islands in de Sitter}
\label{sec:islands-and-inf-par}

\subsection{Is there information paradox in de Sitter?}

The formulation of the information loss paradox in de Sitter can be given in complete analogy with that in the context of black holes. We say about the information loss paradox when the entanglement entropy of thermal radiation coming from the cosmological horizon exceeds the coarse-grained entropy of this horizon, described by the Bekenstein-Hawking formula and called the Gibbons-Hawking entropy $S_\text{GH}$, due to constant growth over time, see~\cite{HIM, Almheiri:2019yqk},
\be
    S\m \gtrsim S_\text{GH} = \frac{\text{Area$\,$(cosmological horizon)}}{4\LP^2},
    \label{eq:S_GH}
\ee
where $\LP = \sqrt{\GN}$ is the Planck length.

\skipline

We have already seen that for the entangling region shown in figure~\ref{fig:ComplPropStatic}, the entanglement entropy is constant. Therefore, it is sufficient to check that this entropy does not exceed the Gibbons-Hawking entropy to establish the absence of the information paradox. When considering this region, we fixed its endpoints on the same hypersurface of constant time. If we relax this requirement, the worst we can get is increase in the entanglement entropy at early times. However, at late times, it reaches a constant value.

The latter statement can be argued as follows. The entropy for the finite region $[\bb_+, \mathbf{0_+}]$ is given by
\be
    \begin{aligned}
        S\m([\bb_+, \mathbf{0_+}]) & = \frac{c}{6}\log\frac{d^2(\bb_+, \mathbf{0_+})}{\eps^2} = \\
        & = \frac{c}{6}\log\left( \frac{2\sqrt{f(b)}}{\kappa^2_c \eps^2} \left[\cosh\kappa_c r_*(b) - \cosh\kappa_c (t_b - t_{0_+})\right]\right).
    \end{aligned}
    \label{eq:Inf-paradox-schematic}
\ee
If $\kappa_c(t_b - t_{0_+}) = \const \neq 0$, the entropy is also constant, as if $t_b = t_{0_+}$. Now, let us assume that $\kappa_c(t_b - t_{0_+}) \equiv g(t_b) \neq \const$. Consider the following function
\be
    \mathcal{F}(t_b) = C - \cosh g(t_b), \quad C = \const. 
\ee
The entropy~\eqref{eq:Inf-paradox-schematic} as a function of $t_b$ can be then represented as
\be
    S\m(t_b) \propto \log \mathcal{F}(t_b), \quad S^\prime(t_b) = \frac{\mathcal{F}^\prime(t_b)}{\mathcal{F}(t_b)}.
\ee
The derivative of $\mathcal{F}^\prime(t_b)$ is given by
\be
    \mathcal{F}^\prime(t_b) = -g^\prime(t_b)\sinh g(t_b).
\ee
Hence, $S\m(t_b)$ monotonically increases if $g(t_b)$ monotonically decreases.

The case, in which the function $g(t_b)$ grows, would cause the Cauchy surface breaking, which is defined as that the initial Cauchy surface, on which the Hilbert space was defined, deforms during time evolution in such a way that it starts to contain timelike subregions. After the Cauchy surface breaking, the problem of calculating the entanglement entropy of a state is no longer well-defined. The moment $\tbreak$, at which this happens, can be found from the equation ${\mathcal{F}(t_b) = 0}$, i.e. when the entropy becomes singular (see~\cite{Ageev:2022qxv} for details). This moment depends on $b$ as
\be
    \tbreak = t_{0_+} + \ell\arctanh\frac{b}{\ell}.
\ee

In the case without the Cauchy surface breaking, the only possibility is that $g(t_b)$ decreases (since we assume that $g(t_b) \neq \const$). This corresponds to the endpoints $\bb_+$ and $\mathbf{0_+}$ approaching the hypersurface of constant time at large times. Therefore, at large times, the function $g(t_b)$ will become small, $g(t_b) \ll 1$, and
\be
    \mathcal{F}'(t_b) \simeq - g'(t_b) g(t_b) \ll 1.
\ee
This allows to conclude that at late times, the entropy for the region $[\bb_+, \mathbf{0_+}]$ saturates at the constant value, given by~\eqref{eq:Inf-paradox-schematic} with $t_b = t_{0_+}$.

\skipline

If there is an island solution for this configuration, it contributes non-trivially only when there is an unlimited monotonic growth in the entanglement entropy. This is the case only for the complement $[\mathbf{0_-}, \bb_+]$ with fixed $t_{0_-}$ (exactly as in the calculation made in~\cite{Sybesma:2020fxg}). However, the island does not heal the violation of complementarity, because the area term is classical and always larger then the semi-classical correction given by the constant entropy for the region $[\mathbf{0_+}, \bb_+]$ in the bulk.


\subsection{The closer to the horizon, the greater the entropy}

\begin{figure}[ht]\centering
    \includegraphics[width=0.85\textwidth]{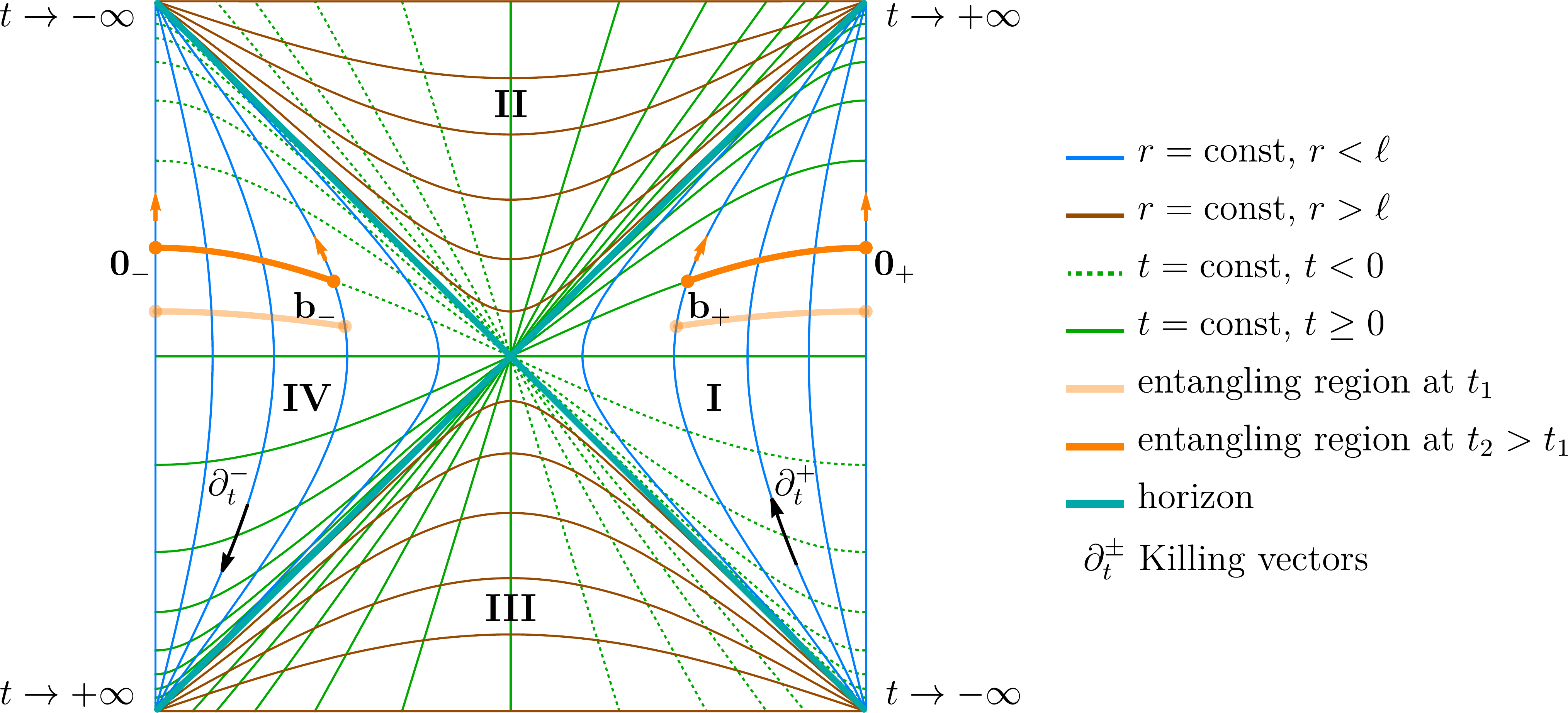}
    \caption{Penrose diagram for the four-dimensional de Sitter spacetime with entangling region ${R = [\mathbf{0_-},\,\bb_-] \cup [\bb_+,\,\mathbf{0_+}]}$, $t_{b_+} = -t_{b_-} = t_{0_+} = t_{0_-}$. Arrows indicate the direction, in which the points flow during time evolution. In the left wedge, the points move in the opposite direction with respect to the Killing vector $\partial^-_t$.}
    \label{fig:FinReg}
\end{figure}

It is reasonable to consider a more complicated configuration, inspired by~\cite{Almheiri:2019yqk, HIM}, which consists of two separate finite entangling regions in the right and left wedges, see~figure~\ref{fig:FinReg}. Let us denote the union of these regions as $R = [\mathbf{0_-},\,\bb_-] \cup [\bb_+\,\mathbf{0_+}]$ with $t_{b_+} = -t_{b_-} = t_b$, $t_{0_+} = t_{b_+}$, $t_{0_-} = t_{b_-}$. By imposing this condition, we get the points of the right wedge moving along the timelike Killing vector $\partial^+_t$ (which is directed from bottom to top of the Penrose diagram), and the points of the left wedge moving \emph{against} the Killing vector $\partial^-_t$ (which is directed from top to bottom). Therefore, this choice of times explicitly breaks the isometry of the problem and makes the setup time-dependent~\cite{Almheiri:2019yqk}.

\begin{figure}[ht]\centering
    \includegraphics[width=0.8\textwidth]{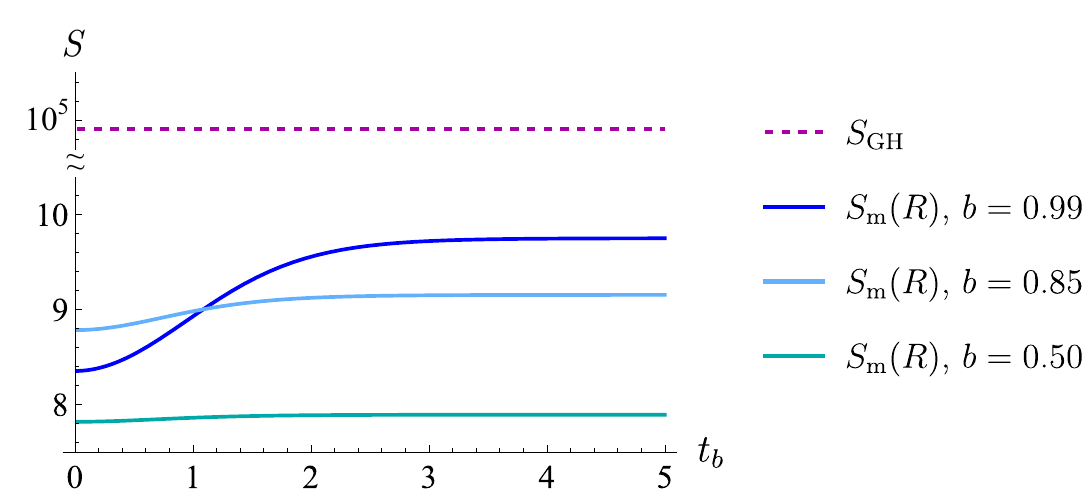}
    \caption{Time evolution of entanglement entropies for region ${R = [\mathbf{0_-},\,\bb_-] \cup [\bb_+,\,\mathbf{0_+}]}$, $t_{b_+} = -t_{b_-} = t_{0_+} = t_{0_-} = t_b$ and different values of $b$. Other parameters are fixed as $\ell = 1$, $c = 3$, $\eps = 0.01$. For any $b$ that is not infinitely close to $\ell$, the entanglement entropy is always less than the Gibbons-Hawking entropy~$S_\text{GH}$ (dashed magenta line).}
    \label{fig:HIMDiffb}
\end{figure}

At early times, such that $t_b \ll r_*(b)/2$, the entropy monotonically increases
\be
    S\m\left(R\right)\EarlyTimes \simeq \frac{2c}{3}\kappa_c t_b + \const.
    \label{eq:linear-growth}
\ee
However, at late times $t_b \gg r_*(b)/2$, the entropy saturates at the value
\be
    S\m\left(R\right)\LateTimes \simeq \frac{c}{3}\log\left(\frac{2\sqrt{f(b)}}{\kappa_c^2\eps^2}\right) + \frac{c}{3}\log\left(\cosh\kappa_c r_{*}(b) - 1\right).
    \label{eq:S-late-times-const}
\ee
Since $\lim\limits_{b\,\to\,\ell} r_*(b) \to \infty$, we conclude that the closer the boundary $b$ of the entangling region to the horizon~$\ell$, the longer the linear growth regime~\eqref{eq:linear-growth}, and the later the entropy reaches saturation, see figure~\ref{fig:HIMDiffb}.

Having said this, we should also emphasize that even though it is formally possible to achieve an excess of entanglement entropy over the Gibbons-Hawking entropy, this seems unlikely in the context of entangling regions located in the de Sitter bulk. By letting the Planck length $\LP = \sqrt{\GN}$ to be equal to the UV cutoff scale~$\eps$ (i.e. we put our system on a lattice with the spacing~$\LP$), we obtain that the Gibbons-Hawking entropy $S_\text{GH}$~\eqref{eq:S_GH} in four dimensions is equal to
\be
    S_\text{GH} = \frac{\pi^2 \ell^2}{\eps^2}.
\ee
This value is much larger than the entropy of matter~\eqref{eq:S-late-times-const}, see also figure~\ref{fig:HIMDiffb}.

\subsection{No island when the paradox can be}

Since the monotonic growth~\eqref{eq:linear-growth} of the entanglement entropy at ${b \to \ell}$ might formally lead to the information paradox, it is interesting to study its behaviour using the island formula~\eqref{eq:Sgen}. In order to find out whether it is possible to avoid the monotonic growth~\eqref{eq:linear-growth} at late times, let us consider a simply connected ansatz for the island
\be
    I = [\mathbf{a_-},\,\mathbf{a_+}].
\ee
This configuration resembles the island for two-sided Schwarzschild black hole~\mbox{\cite{Almheiri:2019yqk, HIM}}. To find the island, it is necessary to study the extremization problem for the generalized entropy functional~\eqref{eq:gen_functional}
\be
    S\gen[I, R] = \frac{\operatorname{Area}(\partial I)}{4\GN} + S\m\left([\mathbf{0_-},\,\bb_-] \cup [\bb_+,\,\mathbf{0_+}] \cup [\mathbf{a_-},\,\mathbf{a_+}]\right).
\ee
However, our numerical analysis shows that the extremization equations have no non-trivial solutions (i.e. $I \neq \varnothing$) at any time. Thus, for the region under consideration, the island does not exist, and the monotonic growth~\eqref{eq:linear-growth} of the entanglement entropy at $b \to \ell$ cannot be avoided.


\subsection{Never dominant island when there is no paradox}

Despite the absence of the information paradox for the configurations considered above (and, as a result, the need to resolve it using the island formula), nevertheless, it is possible to find an island solution for some specific setup.

\begin{figure}[ht]\centering
    \includegraphics[width=0.85\textwidth]{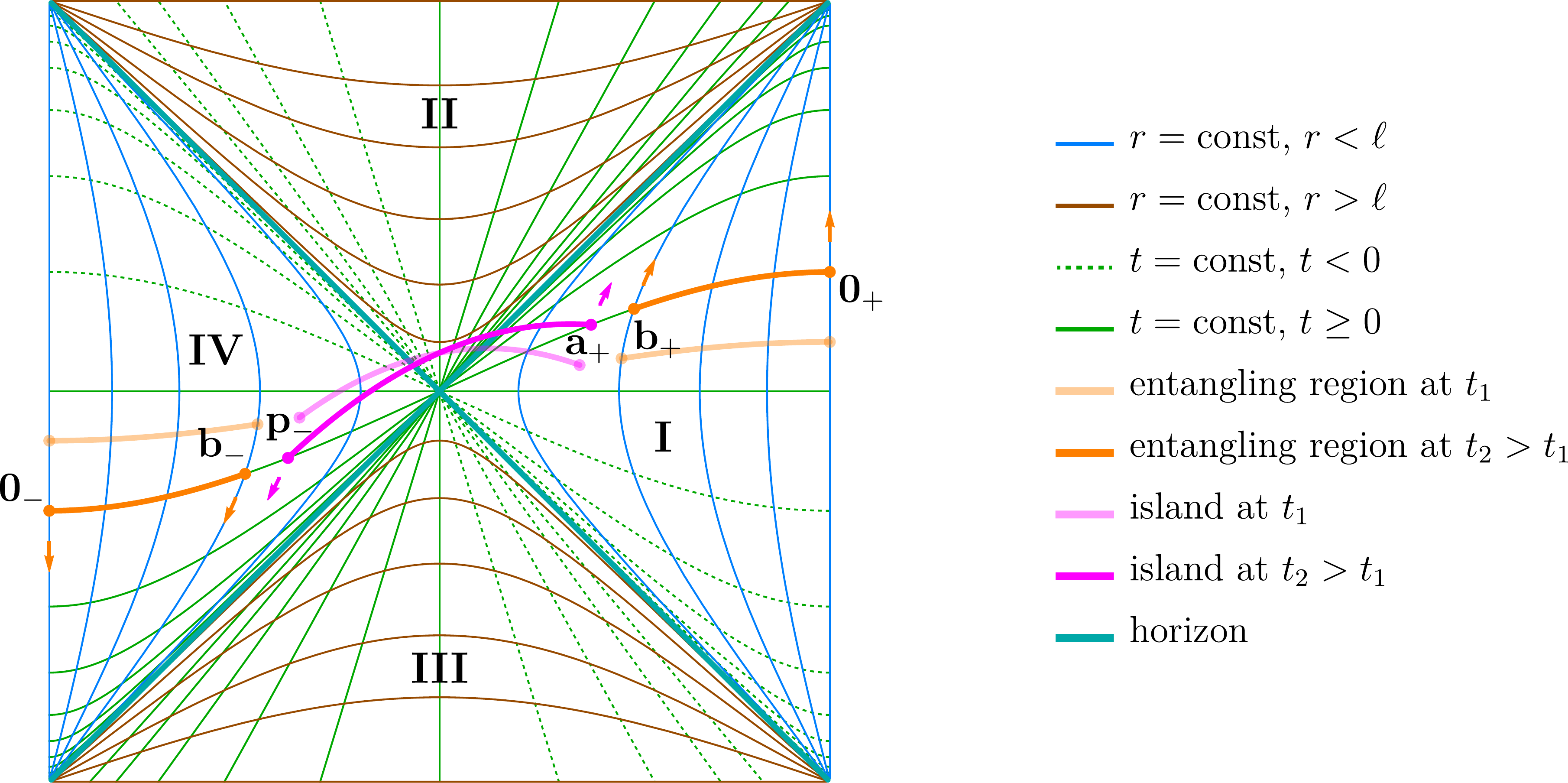}
    \caption{Penrose diagram for the four-dimensional de Sitter spacetime with the entangling region ${R = [\mathbf{0_-},\,\bb_-] \cup [\bb_+,\,\mathbf{0_+}]}$ (orange), $t_{b_+} = t_{b_-} = t_{0_+} = t_{0_-}$. For this configuration, one can find an island, ${I = [\bp_-,\,\ba_+]}$ (magenta). Arrows indicate the direction, in which the points flow during time evolution.}
    \label{fig:FinRegWIsl}
\end{figure}

\begin{figure}[ht]\centering
    \includegraphics[width=0.6\textwidth]{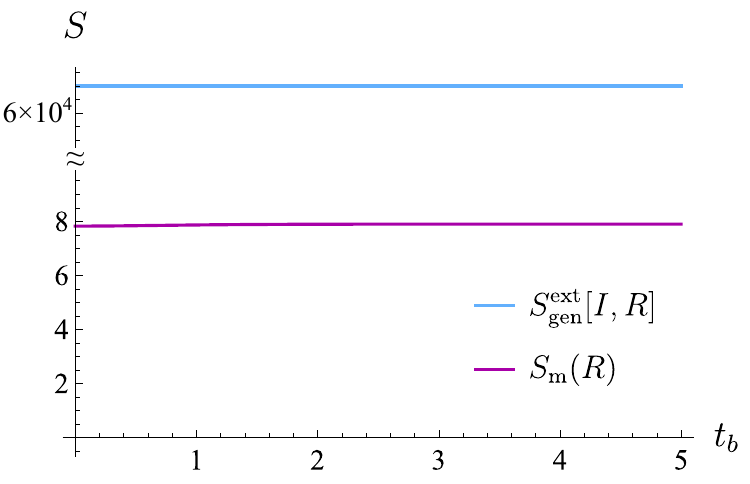}
    \caption{Time evolution of entanglement entropy for entangling region ${R = [\mathbf{0_-},\,\bb_-] \cup [\bb_+,\,\mathbf{0_+}]}$ with $t_{b_+} = t_{b_-} = t_{0_+} = t_{0_-} = t_b$. The parameters are fixed as $b = 0.5$, $\ell = 1$, $c = 3$, $\GN = 10^{-4}$, $\eps = 0.01$. The island contribution $S\extgen$ never dominates since it is always larger than the entropy for the configuration without island $S\m$.}
    \label{fig:NeverDomIsland}
\end{figure}

Let us consider the following configuration (see figure~\ref{fig:FinRegWIsl})
\be
    R = [\mathbf{0_-},\,\bb_-] \cup [\bb_+,\,\mathbf{0_+}], \quad t_{b_+} = t_{b_-} = t_{0_+} = t_{0_-}.
\ee
All the endpoints lie on the same time slice, hence, the entropy for the region~$R$ is constant in time. Consider the following island ansatz $I = [\bp_-,\,\ba_+]$. The extremization problem for the generalised entropy functional~\eqref{eq:gen_functional} with this ansatz yields a symmetric solution
\be
    a_+ = p_- = a, \quad t_{a_+} = t_{p_-} = t_a.
\ee
Since the endpoints lie on the same time slice, the island solution is also constant. Even though there is no information paradox, we have found the island, which, however, does not dominate the generalized entropy functional during time evolution --- even at late times, see figure~\ref{fig:NeverDomIsland}.

We conclude this section with a remark that since possible island solutions do not dominate or even not emerge at all, they do not change the results of our discussion on pure state condition and complementarity in section~\ref{sec:no-purity}.


\section{Conclusions}
\label{sec:conclusions}

In this paper, we have studied the entanglement entropy of $c$ copies of free massless Dirac fermions in two-dimensional de Sitter obtained by partial dimensional reduction of the four-dimensional de Sitter. Our approximation for entanglement entropy can be additionally motivated by the s-mode approximation for matter.

As a consistency test of this setup, we have studied the basic properties of entanglement entropy for the case of a pure quantum state of the entire system, given on a Cauchy surface in de Sitter. We have considered two types of Cauchy surfaces defined in the extended static coordinates (for example, surfaces of constant static time) and in the flat coordinates (surfaces of constant flat time). For both types, pure state condition (vanishing of the Cauchy surface entanglement entropy) and the complementarity property (the equality of the entanglement entropy of a region and its complement) are studied. The main result is that both pure state condition and complementarity are \emph{violated}, seemingly disproving the applicability of the CFT formulas~\eqref{eq:S_1_ints} and~\eqref{eq:S_N_ints} for calculating entanglement entropy to the partially reduced higher-dimensional de Sitter geometry. One possible explanation would be that the partial reduction, which has been considered here in a manner of previous studies, changes the structure of Cauchy surfaces in de Sitter by adding boundaries, thus requiring formulas of BCFT instead of CFT. We leave this question for future research.

For the case of the \emph{extended static patch}, we have found that the entanglement entropy of a Cauchy surface is bounded from below by a non-zero value, which depends on the radiation temperature of the cosmological horizon and some UV cutoff. The difference between the entanglement entropies of a region and its complement for this type of Cauchy surface essentially depends on how the time coordinates are chosen. These entropies are never equal to each other, and the best we can achieve, by choosing time coordinates in a certain way, is that they differ by a constant. However, there are choices of time coordinates that can cause the difference between the entropies for a region and its complement to increase with time. In particular, this can lead to an essentially incorrect interpretation of time dependence of the entanglement entropy. For example, if the entanglement entropy of a certain region is constant, then the entropy of the complement with a special choice of time coordinates of the south and north poles can monotonically increase with time.

It is interesting to compare the properties of entanglement entropy in two-sided Schwarzschild black hole, which were studied in \cite{Ageev:2022qxv}, with our results for the entropy in the extended static patch of de Sitter. For both spacetimes, pure state condition and complementarity are violated, and the entanglement entropy for a Cauchy surface in two-sided Schwarzschild black hole is bounded from below by the same constant. The essential difference is that for two-sided Schwarzschild black hole, it is necessary to regularize spacelike infinities $i^0$ in the left and right wedges in the particular way, which can be done along some curves in the space of radial and time coordinates of the IR regulators~$\bq_\pm$. However, there is no functional dependence of the entanglement entropy on the time coordinates of the regulators after taking the limit $\bq_\pm \to i^0$ to derive the entropy of the full Cauchy surface. In de Sitter, the boundary points of a Cauchy surface in the extended static patch have \emph{finite} spatial coordinates. This fact causes a significant dependence of the entropy for the Cauchy surface on the time coordinates of the poles.

For Cauchy surfaces in \emph{flat coordinates}, the violation of pure state condition and complementarity is even more serious. Verification of these properties requires the regularization of the spacelike infinity $\rho \to \infty$. We have shown that the entanglement entropy for a regularized Cauchy surface and the difference between the entropies for a finite region and its semi-infinite complement diverge in the limit $\rho \to \infty$. We believe that it is impossible to regularize the spacelike infinity in de Sitter in such a way that all the properties of entanglement entropy are satisfied.

We also explore the information paradox in de Sitter and how it can be resolved using the island formula for entanglement entropy. We give an example of a region in the extended static patch, reminiscent of the setup studied for two-sided Schwarzschild black hole, for which there is \emph{time-limited monotonic growth} that could potentially lead to the information paradox, and investigate whether it is possible to find the entanglement island in this case. Numerical calculations show that there is no non-trivial island configuration for this setup, so the possible information paradox \emph{cannot} be resolved by the island formula. We also give an example of a region in the extended static patch, for which there is a \emph{non-trivial} island configuration not previously described in the literature. However, the entanglement entropy of this region does not increase with time, hence, does not lead to the information paradox.


\section*{Acknowledgements}

We would like to thank A.V. Ermakov for helpful discussions. DA, IA and TR are supported by the Russian Science Foundation (project 20--12--00200, Steklov Mathematical Institute). DA, AB and VP are supported by the Foundation for the Advancement of Theoretical Physics and Mathematics ``BASIS''.


\bibliography{IslandsdS}
\bibliographystyle{JHEP}


\end{document}